\documentclass[%
 aip,
 jmp,%
 amsmath,amssymb,
preprint,%
]{revtex4-1}

\usepackage{graphicx}
\usepackage{dcolumn}
\usepackage{bm}
\usepackage{textcomp}
\usepackage{gensymb}
\usepackage{floatrow}
\usepackage{hyperref}
\usepackage{xcolor}
\hypersetup{
    colorlinks=true,
    linkcolor=blue,
    citecolor=blue,
    urlcolor=blue,
}
\usepackage{color,soul}

\begin{document}
\preprint{Sheng2019}
\title{Magnetodynamics in orthogonal nanocontact spin-torque nano-oscillators based on magnetic tunnel junctions}

\author{Sheng Jiang}
\affiliation{Department of Applied Physics, School of Engineering Sciences, KTH Royal Institute of Technology, Electrum 229, SE-16440 Kista, Sweden}
\affiliation
{Department of Physics, University of Gothenburg, 412 96, Gothenburg, Sweden}


\author{Martina Ahlberg}
\affiliation{Department of Physics, University of Gothenburg, 412 96, Gothenburg, Sweden}

\author{Sunjae Chung}
\affiliation
{Department of Applied Physics, School of Engineering Sciences, KTH Royal Institute of Technology, Electrum 229, SE-16440 Kista, Sweden}

\author{Afshin Houshang}
\affiliation{Department of Physics, University of Gothenburg, 412 96, Gothenburg, Sweden}
\affiliation{NanOsc AB, Kista 164 40, Sweden}

\author{Ricardo Ferreira}
\affiliation{International Iberian Nanotechnology Laboratory (INL), Av. Mestre Jos\'e Veiga s/n, 4715-330, Braga, Portugal}
\author{Paulo Freitas}
\affiliation{International Iberian Nanotechnology Laboratory (INL), Av. Mestre Jos\'e Veiga s/n, 4715-330, Braga, Portugal}

\author{Johan~\AA{}kerman}
 \altaffiliation[Correspondence to ]{johan.akerman@physics.gu.se }
\affiliation
{Department of Applied Physics, School of Engineering Sciences, KTH Royal Institute of Technology, Electrum 229, SE-16440 Kista, Sweden}

\affiliation
{Department of Physics, University of Gothenburg, 412 96, Gothenburg, Sweden}
\email{johan.akerman@physics.gu.se }
\affiliation{NanOsc AB, Kista 164 40, Sweden}

\date{\today}

\begin{abstract}
We demonstrate field and current controlled magnetodynamics in nanocontact spin-torque nano-oscillators (STNOs) based on orthogonal magnetic tunnel junctions (MTJs). We systematically analyze the microwave properties (frequency $f$, linewidth $\Delta f$, power $P$, and frequency tunability $df/dI$)  with their physical origins---perpendicular magnetic anisotropy (PMA), damping-like and field-like spin transfer torque (STT), and voltage-controlled magnetic anisotropy (VCMA). These devices present several advantageous characteristics:  high emission frequencies ($f>  20$ GHz), high frequency tunability ($df/dI=0.25$~GHz/mA), and zero-field operation ($f\sim 4$ GHz). Furthermore, a detailed investigation of $f(H, I)$ reveals that $df/dI$ is mostly governed by the large VCMA (287~fJ/(V$\cdot$m)), while STT plays a negligible role.

\end{abstract}

\pacs{Valid PACS appear here}
       
\maketitle



Magnetic tunnel junctions (MTJs)\cite{Moodera1995,Miyazaki1995,Bowen2001,Faure-Vincent2002} are multilayer stacks consisting of two ferromagnetic (FM) layers separated by an insulating tunnel barrier. The advantage of MTJs is their high tunnel magnetoresistance (TMR),\cite{Mathon2001,Bowen2001,Faure-Vincent2002,Ikeda2008b} which enables spintronic applications such as spin-transfer torque magnetoresistive random access memory (STT-MRAM),\cite{Akerman2005,Engel2005,Chappert2007,Bhatti2017MatTod,Locatelli2014} spin-torque nano-oscillators (STNOs),\cite{SIKiselev, Rippard2004,Chen2016, Goto2018,Tarequzzaman2019} 
and neuromorphic computing.\cite{Torrejon2017, Romera2018Nature} Recently, voltage-controlled magnetic anisotropy (VCMA)\cite{Petit2007a,Kubota2007,Li2008,Deac2008,Kalitsov2009,Wang2009} was demonstrated in MTJ structures, which is very promising for next-generation devices, including highly energy-efficient magnetoelectric RAM (MeRAM).\cite{Amiri2015,Wang2017} 

In ordinary STNOs, high current densities are needed to provide sufficient spin torque to counteract the intrinsic damping and to maintain steady-state auto-oscillations. Such high current densities can be achieved either by patterning the entire multilayer to a nanopillar (NP) or by confining the current flow path via a nanocontact (NC) on top of the stack. 
The nanopillar approach gives a lower threshold current and higher emission power,\cite{Deac2008,Maehara2013a} but also results in a larger linewidth than with NC-STNOs. The major benefits of nanocontact devices are their high-frequency operation, the wide range of different distinct spin wave modes, and mutual synchronization.\cite{Mancoff2005a,Kaka2005a,Sani2013,Houshang2016,Houshang2018NC} The  MTJ structure's major issue of current shunting through the cap layer can be solved by fabricating so-called sombrero-shaped NC MTJ-STNOs.\cite{Maehara2013a,Maehara2014,Houshang2018NC} Such devices with in-plane (IP) magnetized layers have been extensively studied,\cite{Deac2008,Muduli2011b,Maehara2014,Romera2015,Sharma2017a,Houshang2018NC} 
while the nanocontact geometry for MTJs with strong perpendicular anisotropy (PMA) remains largely unexplored, even though it should offer the potential of zero-field operation\cite{Fowley2014} and high frequency-tunability\cite{Goto2018} due to strong PMA.
In addition, nano-contact STNOs with PMA free layers can support magnetic droplet solitons,\cite{Mohseni2013,Macia2014,Chung2016,Chung2018PRL} although these have not yet been demonstrated in MTJ based devices.

In this letter, we investigate the magnetodynamics of orthogonal MTJ-STNOs with a sombrero-shaped nanocontact, using a wide range of fields (up to 0.65~T) and currents (up to $\pm $12~mA). Measurements reveal a clear 4~GHz microwave signal at zero applied field, and the frequency of the main peak follows  typical FMR-like behavior with increasing field. The current dependence of the frequency can be parametrized using a second-order polynomial, which captures the effects of the Joule heating, the Oersted field, the field-like spin-torque, the VCMA, and the magnetic field. We show that the VCMA is the dominating term that governs the linear contributions to the frequency tunability, while the field-like torque plays only a minor role. No evidence for magnetic droplet solitons were observed. The results provide important contributions to the development of current-tunable MTJ based STNOs with a broad frequency range and low-field operation.

\begin{figure}
\centering
\includegraphics[width=3.2in]{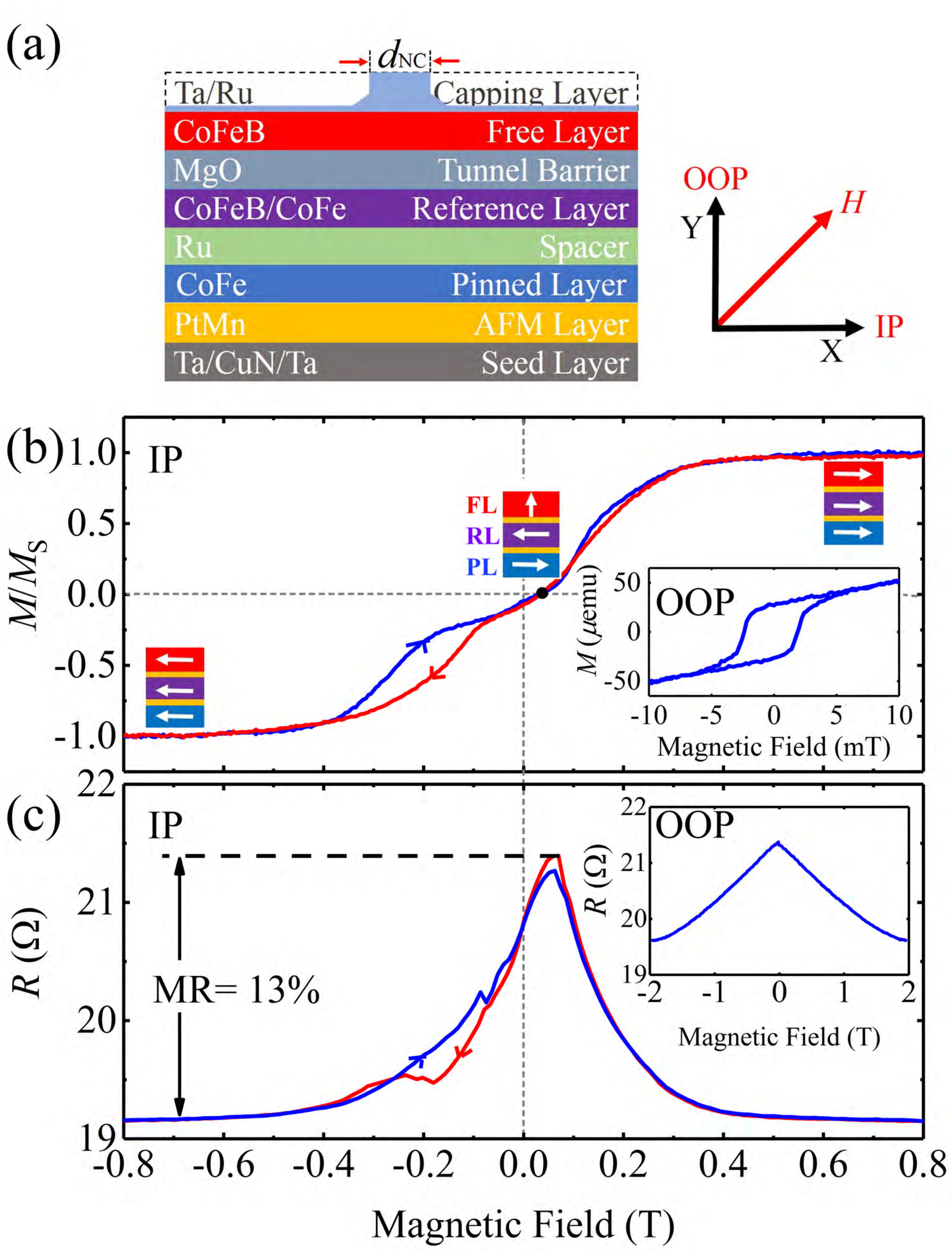}
\caption{\label{fig1} (a) Schematic of a ``sombrero" MTJ-STNO, together with the direction of in-plane (IP) and out-of-plane (OOP) magnetic fields. (b) Hysteresis loop of an unpatterned MTJ stack in an IP field. The magnetic states of the three magnetic layers [free (FL), reference (RL), and pinned layer (PL)] are depicted at three different field positions by the white arrows. The inset shows the minor hysteresis loop in an OOP field.  (c) Resistance of an MTJ-STNO versus the IP magnetic field, revealing an MR of 13\%. The arrows on the lines display the field-sweep directions. The inset presents the resistance as a function of OOP field. }
\end{figure}


The MTJ stack was deposited on a Si/SiO$_{2}$ substrate using magnetron sputtering. The stack, illustrated in Fig.~\ref{fig1}(a), consists of two FM layers (CoFeB and CoFeB/CoFe) separated by an MgO tunneling barrier with a resistance--area (RA) product of about 1.5~\ohm$\cdot \mu$m$^2$. The top CoFeB layer with strong PMA acts as the free layer (FL), while the bottom, easy-plane, CoFeB/CoFe acts as the reference layer (RL). The pinned layer (PL) is made of CoFe, and is separated from the RL by a thin layer of Ru. An antiferromagnetic (AFM) PtMn layer is located immediately below the PL. The seed layer is Ta/CuN/Ta. The full stack is hence SiO$_{2}$ substrate/Ta(3)/CuN(30)/Ta(5)/PtMn(20)/Co$_{70}$Fe$_{30}$(2)/ Ru(0.7)/Co$_{60}$Fe$_{40}$B$_{20}$(2)/Co$_{70}$Fe$_{30}$ (0.5)/MgO(0.7)/ Co$_{60}$Fe$_{40}$B$_{20}$(1.4)/Ta(3)/Ru(7), where the numbers in parentheses are the thicknesses in nanometers. The stack was then processed into nanocontact spin-torque nano-oscillators with a 125-nm nominal raidus as described in Ref.~[\onlinecite{Houshang2018NC}]. To force a higher current through the MgO tunneling barrier, the capping layer was gradually thinned laterally by ion milling, \cite{Maehara2013a,Maehara2014,Houshang2018NC} as shown in Fig.~\ref{fig1}(a). 
Magnetic hysteresis loop measurements were conducted using an alternating gradient magnetometer (AGM). For dc and microwave characterization, we used a custom-built 40 GHz probe station; the details of this setup can be found in Ref.~[\onlinecite{Jiang2018PRA}].


\begin{figure}
\centering
\includegraphics[width=2.8in]{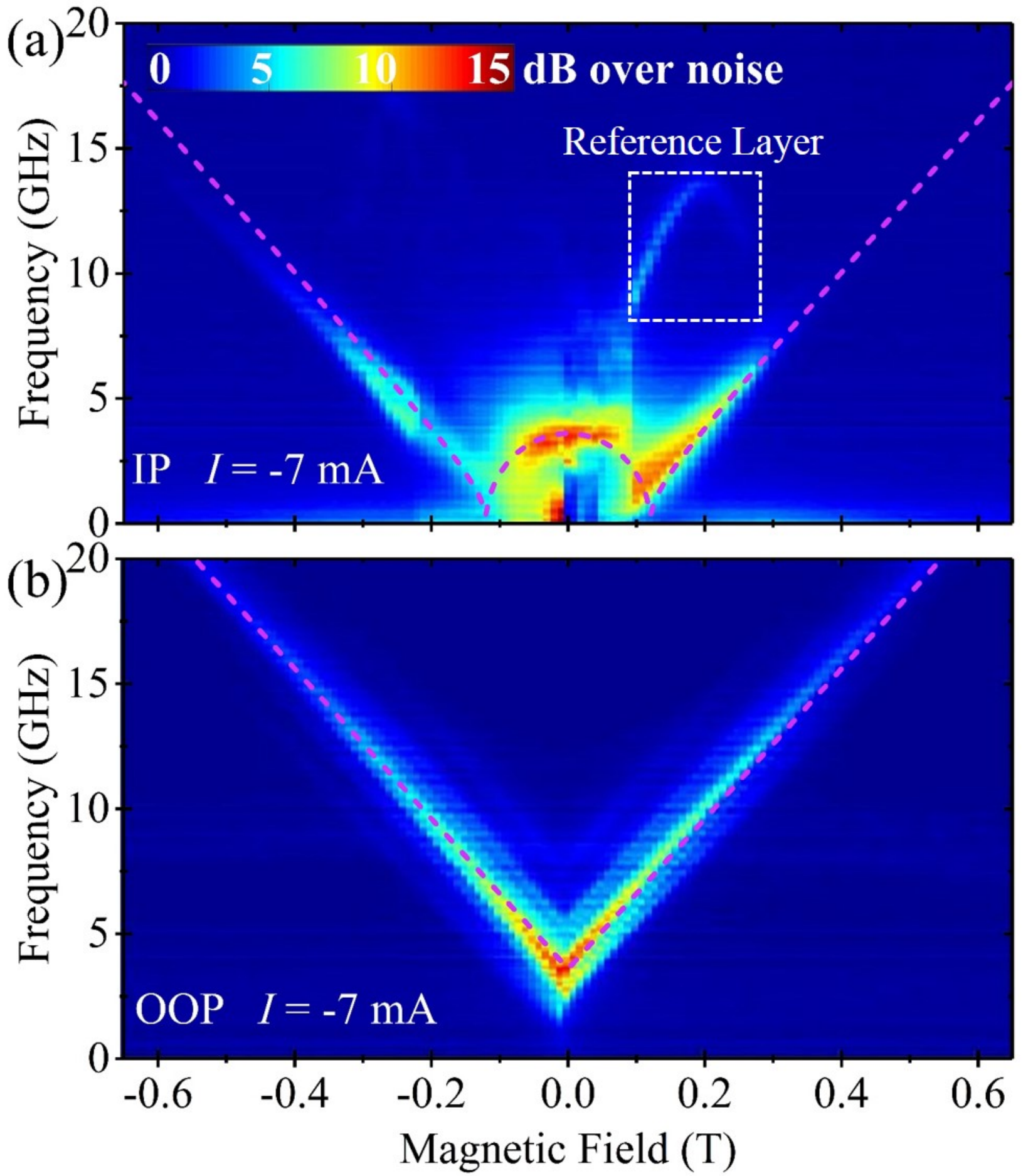}
\caption{\label{fig2} Power spectral density versus (a) IP and (b) OOP field at $I=-7$~mA. The nominal nanocontact radius is 125 nm. The dashed lines are the calculated FMR frequencies for IP and OOP fields, using a nominal $\mu_{0} M_{\text{eff}}=-0.12$~T.
}
\end{figure}

The in-plane (IP) hysteresis loop and the magnetoresistance curve are presented in Figs.~\ref{fig1}(b) and (c), respectively. The corresponding out-of-plane (OOP) responses are shown in the insets. The orthogonal magnetic configuration is evident and the magnetoresistance is 13\%. 

Figs.~\ref{fig2}(a) and (b) show the generated power spectral density (PSD) at $I_{\text{dc}} = -7$~mA, in IP and OOP  fields, respectively. The main peak reflects the ordinary auto-oscillation of the free layer, which follows the Kittel equation:
\begin{equation}
\label{eq1}
f_{\mathrm{FMR}}=\left\{\begin{array}{ll}{\frac{\gamma \mu_{0}}{2 \pi} \sqrt{H_{\mathrm{res}}\left( \left| H_{\mathrm{res}}+M_{\mathrm{eff}} \right| \right)},} & {\mathrm{IP}} \\ {\frac{\gamma \mu_{0}}{2 \pi}\left( \left| H_{\mathrm{res}}-M_{\mathrm{eff}} \right| \right),} & {\mathrm{OOP}}\end{array}\right\},
\end{equation}
\noindent where $\gamma /2\pi=31$~GHz/T is the gyromagnetic ratio, $\mu _{0}$ is the permeability of free space, $H_{\mathrm{res}}$ is the resonance field (\emph{i.e.}~the applied field), and the effective magnetization, $\mu_{0} M_{\text{eff}} =\mu_{0}( M_\mathrm{s}- H_\mathrm{k})= -0.12$~T, has been chosen to give  good agreement between the experimental results and the calculated resonance fields. $H_\mathrm{k}$ is the perpendicular magnetic anisotropy field; the negative value of $M_{\text{eff}}$ indicates that the free layer has an OOP easy axis.

The IP measurements in Fig.~\ref{fig2}(a) show the rich dynamics of the measured spectrum, where not only are the free layer oscillations  detected, but a high-frequency peak from the reference layer\cite{Zeng2010b} appears during the switching of the  layer. This response is highlighted by the  white dashed rectangle in Fig.~\ref{fig2} (a). The details of the collective response are beyond the scope of this study;  we focus instead on the free layer behavior. The OOP field dependence of the frequency is symmetric for the magnetic field, with side peaks accompanying the main signal.  The intensity of the signal decreases with field (in both configurations), since the precession angle of the free layer decreases. We further observe a clear strong peak ($f\approx 4$~GHz) at zero field ($\mu_{0}H=0$~T), which implies that this type of STNO can be used as a field-free high-frequency microwave generator.
\cite{Zhou2009a,Skowronski2012b, Fowley2014, Yamamoto2015a,Fang2016a,Kowalska2019} The zero-field frequency can then be increased if a free layer with stronger PMA is used.

\begin{figure}
\centering
\includegraphics[width=3.37in]{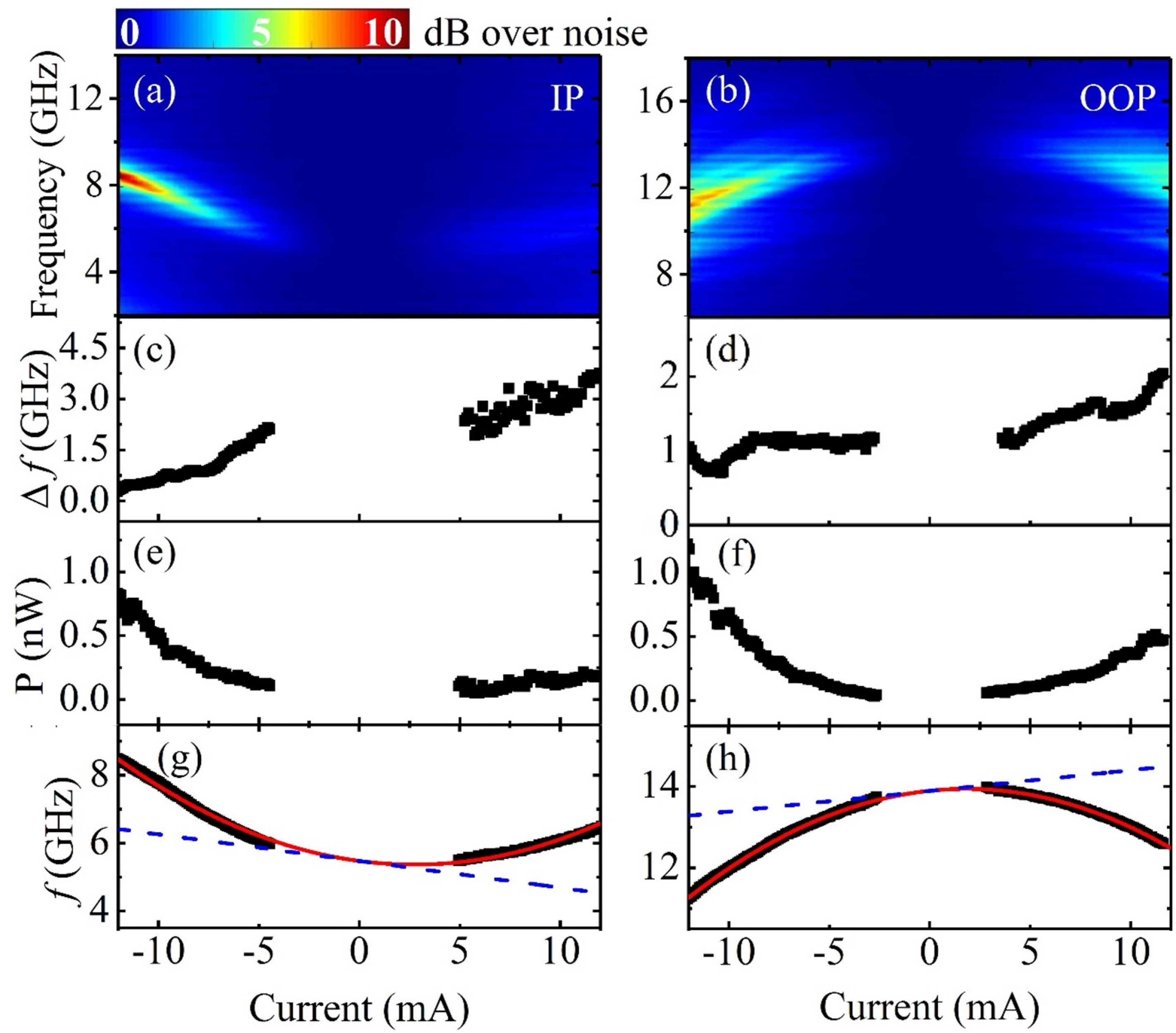}
\caption{\label{fig3} PSD as a function of bias current in an (a) IP or (b) OOP field of $\mu_{0} H = -0.32$~T. The extracted linewidth $\Delta f$, the integrated power $P$, and the frequency $f$ are shown in (c)--(h). The solid red lines show the quadratic polynomials $f = A I^2 + B I + C$ fitted to the data. The blue dashed lines represent the linear and constant terms of the fits ($A=0$).
}
\end{figure}

The frequency $f$, linewidth $\Delta f$, and power $P$ are three key properties of microwave generators. 
The current dependent PSD of the auto-oscillations is presented in Fig.~\ref{fig3}, along with the extracted $f$, $\Delta f$, and $P$. We observe microwave signals at both negative and positive currents up to 12~mA, which shows that both thermal fluctuations and spin-transfer torque (STT) can facilitate the magnetodynamics of a free layer. The effect of STT can be described by the Landau--Lifshitz--Gilbert equation with STT terms,\cite{Zhang2002}
\begin{equation}
\begin{split}
\frac{d\textbf{M}}{dt}=-\gamma \textbf{M}\times \textbf{H}_{\text{eff}}+\frac{\alpha }{M_{\text{s}}}\textbf{M}\times \frac{d\textbf{M}}{dt}\\-\frac{\gamma a_{\text{j}}}{M_{\text{s}}}\textbf{M}\times (\textbf{M}\times \textbf{P})-\gamma b_{\text{j}}\textbf{M}\times \textbf{P},
\end{split}
\end{equation}
where $\alpha $ is the intrinsic damping constant, $M_{\text{s}}$ is the saturation magnetization of the free layer, $H_{\text{eff}} = H + H_{\text{k}}- M_{\text{s}}$ is the effective field, \textbf{M} is the magnetization vector of the free layer, and \textbf{P} is the polarization vector of the spin current (related to the reference layer's magnetization). The two terms on the second row of the equation are the Slonczewski in-plane STT and the perpendicular field-like torque (FLT), respectively. Both $a_{\text{j}}$ and $b_{\text{j}}$ are proportional to the dc current density $j$.\cite{Petit2007a} Note that $b_{\text{j}}$ is negligible in metallic spin-valves, while it is comparable to $a_{\text{j}}$ in MTJs.\cite{Sankey2007,Petit2007a,Li2008}  

Figures~\ref{fig3}(c) and (d) shows $\Delta f$ as function of the applied current. Generally, it shows a linearly proportional behavior as the  current increases for both IP and OOP fields. For the IP field, the linewidth increases linearly with increasing positive current, while it decreases linearly for negative currents. This observation can be explained by the damping-like torque provided by the STT, which enlarges or reduces the effective $\alpha$, depending on the polarity of the current ($+/-$). As this effect only matters to the current, regardless of the applied field direction, we indeed observe similar behavior again for the OOP field in Fig.~\ref{fig3}(d). 

The generated integrated power ($P$) was calculated from Lorentz fitting of the spectra after correcting for the impedance mismatce; $P$ is plotted in Figs.~\ref{fig3}(e) and (f). Again, the overall trends are similar for IP and OOP fields. $P$ is quadratic in $I$ and increases more rapidly for negative currents than for  positive. This asymmetry arises because the STT either facilitates or impedes oscillations, depending on the polarity, while the Joule heating assists the magnetodynamics equally for the currents of both polarities.

The frequency ($f$) is well fitted by a second-order polynomial, $f = A I^2 + B I + C$, which is plotted as the red lines in Figs.~\ref{fig3}(g) and (h). The first term ($A I^2 $) is symmetric around zero and originates from  Joule heating,\cite{Petit-Watelot2012c} which causes temperature-induced modifications of $M_\text{s}$ and $H_\text{k}$.\cite{Guo2015a,Wang2017T} Consequently, the frequency increases (decreases) for IP (OOP) fields as $M_\text{eff}$ varies with current and temperature. Another possible contribution to the coefficient $A$ is the Oersted field,\cite{Arias2007,Consolo2007,Jiang2018Oe} which is perpendicular to, and thus independent of, the current direction.

\begin{figure}
\centering
\includegraphics[width=3.4in]{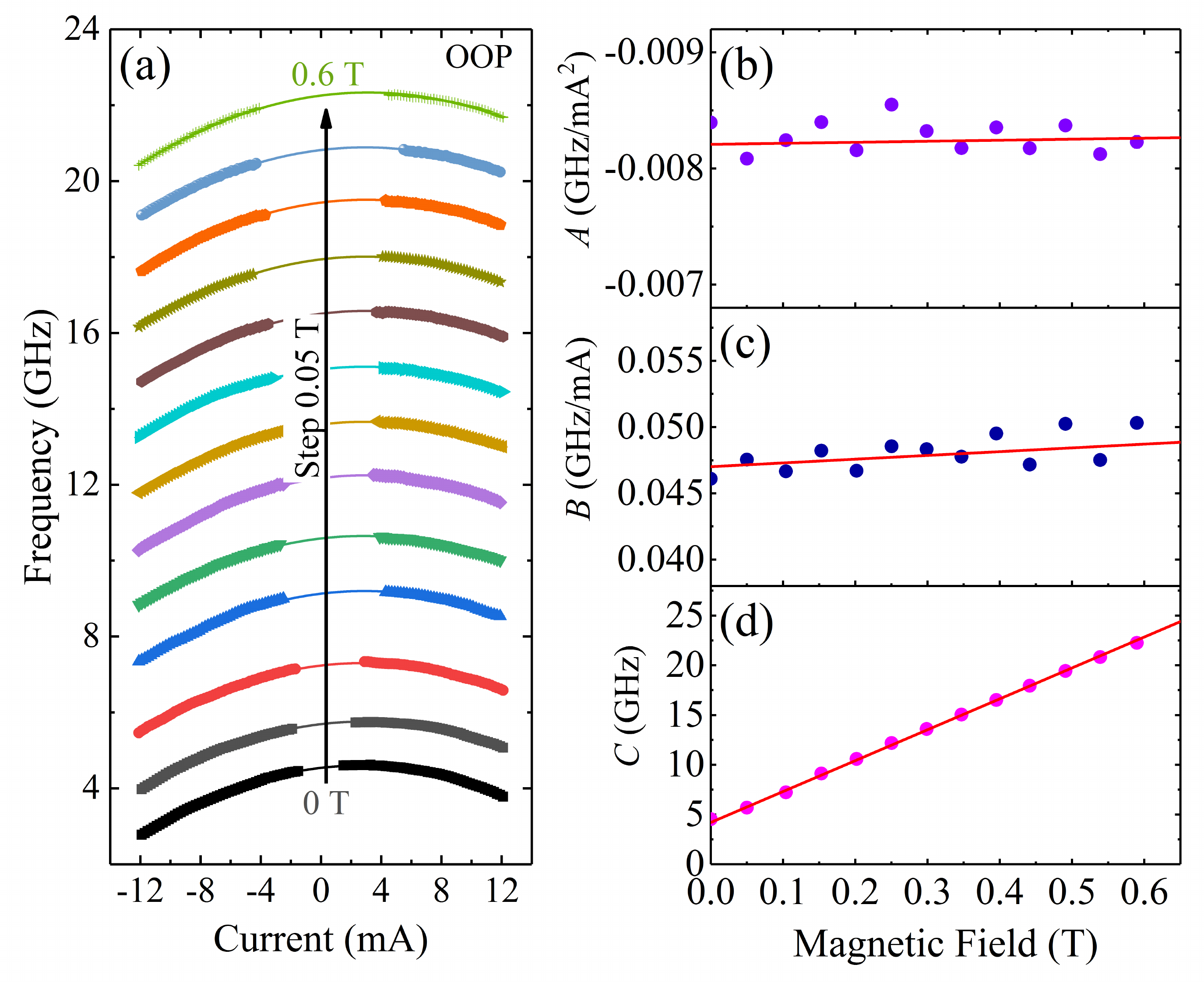}
\caption{\label{fig4} (a) Frequency as a function of current at different OOP fields. The data is fitted by $f = A I^2 + B I + C$, illustrated by the solid lines. (b--d) The extracted coefficients $A$, $B$ and $C$ vs. field together with linear fits. 
}
\end{figure}

The asymmetric term ($B I$) can in principle be ascribed to both spin torque and VCMA. The FMR frequency depends on the current density as $df/dj_\text{dc} \propto -2\alpha a_{\text{j}} / j+b_{\text{j}} /j$,\cite{Petit2007a} where the Slonczewski term must be much smaller than the FLT, given the small damping $\alpha \ll 1 $. Both terms are proportional to $j$, making the frequency change proportional to $I$. The anisotropy is tuned with applied voltage $(U)$,\cite{Petit2007a,Kubota2007,Li2008,Deac2008,Kalitsov2009,Wang2009,AMIRI2012} and this VCMA is linear in $U$.\cite{Petit2007a,Kubota2013a} Thus, the effective magnetization is altered by $U(I)$ and the frequency varies according to Eq.~(1)---that is, the impact is different for the IP and OOP fields. The difference is small but visible in Figs.~\ref{fig3}(g) and (h). The fit deviates slightly from the IP data, but matches the OOP frequency perfectly, which also validates the assumption that $U \propto I$. 


The different contributions to $f \left( I \right)$ can be further elucidated by analyzing the field dependence of the coefficients. Fig.~\ref{fig4}(a) shows current-sweep measurements in OOP fields between $0$ and $0.6$~T. The data is well described by second-order polynomials, and the results of the fits are presented in Figs.~\ref{fig4}(b)--(d). The coefficient $A$ is independent of field, as expected for Oersted-field and Joule-heating effects in Fig.~\ref{fig4}(b). The offset $C$ corresponds to the zero-current frequency and follows the Kittel equation. A fit of Eq.~(\ref{eq1}) to the data in Fig.~\ref{fig4}(d) gives $\mu_{0} M_{\text{eff}} = -0.135$~T. The discrepancy between this result and the fit to the frequencies at $I=-7$~mA (Fig~\ref{fig2}) is likely caused by heating. Note that both values of $\mu_{0} M_{\text{eff}}$ are also predicted to differ from the true value that governs the FMR, due to the inherent nonlinearity of STNOs.\cite{Andrei2009,Jiang2018N}

We now focus on the relative effects  of FLT and VCMA, and we  also examine the coefficient $B$. The FLT strength is a sine function of the angle $\theta $ between $\textbf{M}$ and $\textbf{P}$,\cite{Wang2009} and will vary with applied field. The angle was calculated from the MR curve in the inset of Fig.~\ref{fig1}(c), and $\theta $ changes from about 90\degree ~at $\mu_{0} H=0$ to 36\degree ~at $\mu_{0} H=0.6$~T. This corresponds to a 70\% decline in FLT, which clearly is not reflected by the data in Fig.~\ref{fig4}(c). In contrast, the VCMA should be virtually independent of the field, since it originates with the applied voltage. Hence, the tiny change in $B \left( H \right) $ demonstrates that the VCMA dominates the linear term. The calculated value of the induced PMA shift is approximately 287~fJ/(V$\cdot$m); this  high value is similar to those reported for FeB/MgO,\cite{Goto2018} CrFe/MgO,\cite{Nozaki2017} and Fe/MgO.\cite{Nozaki2016} 

Our straightforward description of the current--field dependence makes possible a simple method for tailoring the frequency tunability ($df/dI=2AI+B$) by selectively choosing $I$ and $H$; we thus obtain, for example, $df/dI=0.25$ GHz/mA at $I=-12$~mA and $\mu_{0} H=0.06$~T. Moreover, by engineering the free-layer material and the interface, it is possible to decide whether the tunability should vary strongly or weakly with $I$. Large temperature effects and small VCMA give a large $A$, while the opposite favors the constant contribution from $B$.


In conclusion, we have studied the magnetodynamics of MTJ-based NC-STNOs with a strong PMA free layer. The current dependence of the frequency can be described using a simple second-degree polynomial, and we have examined the coefficients as a function of field, which allowed us to separate the different mechanisms behind the frequency tunability. Furthermore, we identify three points of importance for applications: i) these STNOs can generate a frequency at 4 GHz even at $H=0$~T, which is essential for the development of field-free microwave generators; ii) the VCMA, which plays a key role in  the linear tunability, has a high value of 287~fJ/(V$\cdot$m); iii) the frequency tunability, which is crucial for the device's operation, can be tailored by selecting the applied field and current, as well as by engineering of the free-layer material.

This work was supported by the China Scholarship Council (CSC), the Swedish Foundation for Strategic Research (SSF), the Swedish Research Council (VR), and the Knut and Alice Wallenberg Foundation (KAW). The work was also partially supported by the European Research Council (ERC) under the European Community's Seventh Framework Programme (FP/2007–2013)/ERC Grant 307144 ``MUSTANG''.


%

\end{document}